\documentclass[sigconf,authorversion]{acmart}

\setcopyright{acmcopyright}
\usepackage{array}
\usepackage{balance}       % to better equalize the last page
\usepackage{graphics}      % for EPS, load graphicx instead 
\usepackage{xcolor}
\usepackage{textcomp}
\usepackage{color}
\usepackage{colortbl}
\usepackage{todonotes}
\usepackage{cuted}
\usepackage{tabularx}
\usepackage{quoting}
\usepackage{multirow}
\usepackage[caption=false, margin=5pt]{subfig}
\usepackage[font={small,bf},labelfont=bf]{caption}
\usepackage{dcolumn}
\quotingsetup{font=itshape,vskip=2pt, leftmargin=2pt}
\newcolumntype{d}[1]{D{.}{.}{#1}}
\newcolumntype{R}{>{\footnotesize}X}

\usepackage{microtype}        % Improved Tracking and Kerning
\usepackage{ccicons}          % Cite your images correctly!
\usepackage{todonotes}

\def\plaintitle{Brotate and Tribike: Designing Smartphone Control for Cycling}
\def\plainkeywords{cycling; smartphone; physical activity; traffic}

\makeatletter
\def\url@leostyle{%
  \@ifundefined{selectfont}{
    \def\UrlFont{\sf}
  }{
    \def\UrlFont{\small\bf\ttfamily}
  }}
\makeatother
\def\pprw{8.5in}
\def\pprh{11in}

\definecolor{linkColor}{RGB}{6,125,233}

\copyrightyear{2020} 
\acmYear{2020} 
\setcopyright{acmcopyright}
\acmConference[MobileHCI '20]{22nd International Conference on Human-Computer Interaction with Mobile Devices and Services}{October 5--8, 2020}{Oldenburg, Germany}
\acmBooktitle{22nd International Conference on Human-Computer Interaction with Mobile Devices and Services (MobileHCI '20), October 5--8, 2020, Oldenburg, Germany}
\acmPrice{15.00}
\acmDOI{10.1145/3379503.3405660}
\acmISBN{978-1-4503-7516-0/20/10}

\begin{document}

\title{\plaintitle}

\author{Pawe\l{} W. Wo\'{z}niak}
\orcid{0000-0003-3670-1813}
\affiliation{
\institution{Utrecht University}
\city{Utrecht}
\country{the Netherlands}}
\email{p.w.wozniak@uu.nl}

\author{Lex Dekker}
\affiliation{
\institution{Utrecht University}
\city{Utrecht}
\country{the Netherlands}}

\author{Francisco Kiss}
\orcid{0000-0003-3670-1813}
\affiliation{
\institution{University of Stuttgart}
\city{Stuttgart}
\country{Germany}}
\email{francisco.kiss@vis.uni-stuttgart.de}

\author{Ella Velner}
\affiliation{
\institution{University of Twente}
\city{Enschede}
\country{the Netherlands}}
\email{p.c.velner@utwente.nl}

\author{Andrea Kuijt}
\affiliation{
\institution{Utrecht University}
\city{Utrecht}
\country{the Netherlands}}

\author{Stella F. Donker}
\affiliation{
\institution{Utrecht University}
\city{Utrecht}
\country{the Netherlands}}
\email{s.f.donker@uu.nl}

\begin{abstract}
The more people commute by bicycle, the higher is the number of cyclists using their smartphones while cycling and compromising traffic safety. We have designed, implemented and evaluated two prototypes for smartphone control devices that do not require the cyclists to remove their hands from the handlebars---the three-button device Tribike and the rotation-controlled Brotate. The devices were the result of a user-centred design process where we identified the key features needed for a on-bike smartphone control device. We evaluated the devices in a biking exercise with 19 participants, where users completed a series of common smartphone tasks. The study showed that Brotate allowed for significantly more lateral control of the bicycle and both devices reduced the cognitive load required to use the smartphone. Our work contributes insights into designing interfaces for cycling.
\end{abstract}

% ACM Classfication


%\printccsdesc

% \section{Stuff needed}
% \begin{itemize}
%     \item Empty Survey
%     \item Survey results
%     \item Photos of low-fi
%     \item Low-fi study results --- comparison + gestures
%     \item Photos of the setup
%     \item Procedure text
%     \item Participant demographics
% \end{itemize}
% \section{Todo}
% \begin{itemize}
%     \item Clean photos of devices
%     \item Video
%     \item Graphs for numbers
%     \item Study figure
% \end{itemize}{}

\maketitle

\begin{teaserfigure}
  \includegraphics[height=4cm]{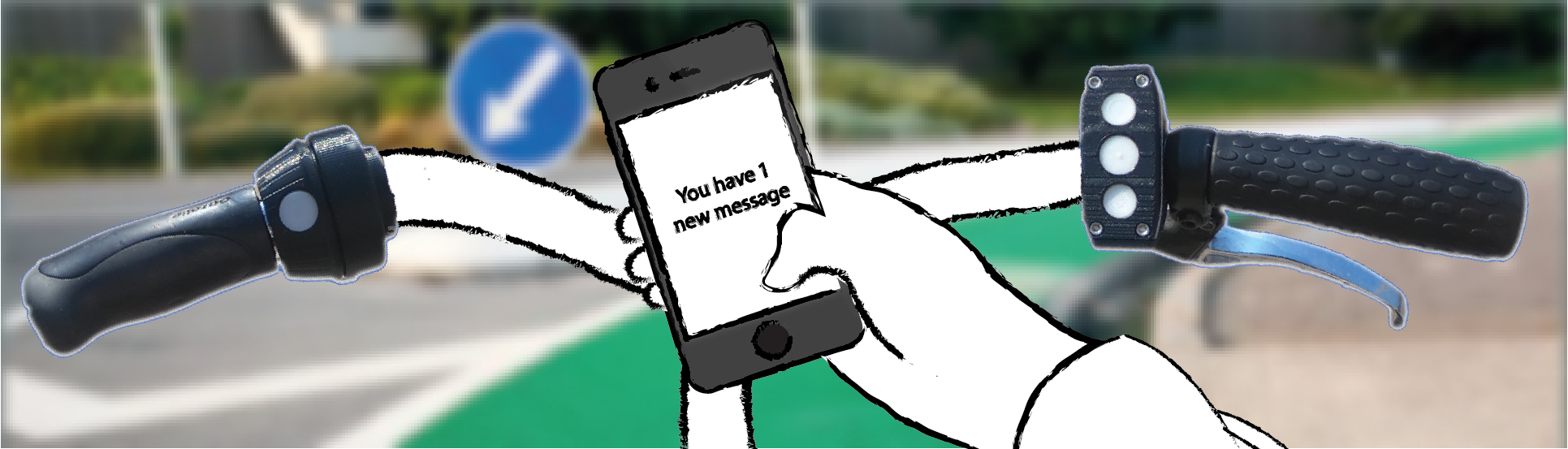}
       \caption{The two controllers for smartphone interaction while cycling---Brotate (left) and Tribike (right)}\label{fig:teaser}
  \label{fig:teaser}
\end{teaserfigure}

\section{Introduction}
All around the world, the number of bike commuters is on the rise~\cite{dill2003bicycle,HARMS2014232}. Research has shown that cycling provides significant health benefits and contributes to more liveable and sustainable cities~\cite{SINGLETON2019249}, and coincides with the rise of smartphones as a primary means of communication~\cite{Fishman,zickuhr2012three}. Studies have shown that cyclists around the globe often listen to music or conduct phone calls while cycling~\cite{GOLDENBELD20121}. Multiple studies~\cite{TERZANO201387,Ichikawa} have shown that cyclists are likely to use their smartphones while in motion, removing their hand from the handlebars and thus increasing the probability of an accident as single-handed cycling significantly reduces lateral control~\cite{DEWAARD2014196}.

As an ever increasing number of  smartphone users cycle and traffic infringement penalties for  using a smartphone while cycling only have limited effects~\cite{Ichikawa}, an alternative approach is needed. Enabling users to use their smartphone while cycling while minimising the decrease in safety for the usage may not only increase traffic safety, but also contribute to further development of cycling as a sustainable means of commuting. Consequently, designing effective smartphone controls for in-ride interactions that minimise distraction and allow for directing more attention to traffic emerges as a challenge for Human-Computer Interaction (HCI).

A number of past works in HCI have addressed interacting with technology on the bicycle, both during and after the ride. Multiple ways of providing input were proposed including gestures~\cite{Dancu:2015:GBE:2817721.2817748}, on-body tapping~\cite{Vechev:2018:MOA:3206505.3206527} and voice interaction~\cite{Soros:2013:CPB:2543651.2543660}, however this work mostly focused on performance in using interaction techniques and did not investigate the impact of the input methods on the control of the bicycle (which has safety implications) or performance in secondary tasks such as smartphone control. In contrast, our work explores the means of performing smartphone tasks while cycling using input methods that do not require removing one's hand from the handlebars, and studies how providing input in-ride affects the stability of the bicycle.

To that end, we conducted a user-centred design process to design alternative handlebar devices for controlling the smartphone while cycling. We first conducted a survey to determine which smartphone functions were most commonly used during cycling. The next step was a pre-study with low-fidelity prototypes to determine the two most promising designs. We then built functional prototypes of Brotate---a smartphone controller where actions are performed by rotating the handlebar grip around an axis parallel to the handlebars, and Tribike---a device with three buttons placed directly next to the handlebar grip. We then compared the two prototypes against using the smartphone held in one hand in an experiment with $n=19$ participants. We contribute: (1) the first, to our knowledge, systematic study of input techniques for smartphone controls during cycling; (2) a generative design contribution of two examples of eyes-free smartphone controls for bicycles developed in a user-centred design process and (3) experimentally determining the desirable properties for future improved smartphone controls for cycling.

\section{Related work}
In this section, we first review past work on interaction for cyclists. We then explore similar designs from the automotive domain and describe our inspirations from work on providing input during physical activity.
\subsection{Interaction for Cyclists}
Designing interactive artefacts and services for cyclists is a recurring topic in HCI research. A significant body of work is dedicated to providing effective output for cyclists, especially for the purposes of navigation. Poppinga et al.~\cite{Poppinga:2009:TTD:1613858.1613911,Pielot:2012:TSE:2371574.2371631} built a tactile display for communicating directional cues. They found that users were able to process signals communicated through vibration motors on the handlebars with medium accuracy. Huxtable et al~\cite{Huxtable:2014:ZBN:2559206.2579481} proposed moving the vibration motors to a user's wrists, and Steltenpohl et al.’s Vibrobelt~\cite{Steltenpohl:2013:VTN:2449396.2449450} was a device that provided output through vibration on the waist. Using Vibrobelt led to fewer navigation errors than a smartphone mounted on the handlebars. Another approach to navigation used street surface for display as in Smart Flashlight~\cite{Dancu:2014:SFM:2556288.2557289}. Matviienko et al.~\cite{Matviienko:2019:NCU:3290605.3300850} found that auditory feedback was preferred by child cyclists. The variety of systems developed for cyclists in past research reflects a demand for the ability to effectively interact with information while cycling, which our work explores. Further, the research also includes a range of design possibilities for modifying existing bike equipment, which inspired our design.

Another strain of HCI work investigates the use of interactive technologies for improving cycling safety. Carton~\cite{Carton:2012:DCA:2370216.2370341} proposed a smart glove for additional safety in indicating directions, resulting in additional visibility. Gesture Bike~\cite{Dancu:2015:GBE:2817721.2817748} implemented gesture recognition to support the automatic activation of turn signals for bicycles. The authors found that users preferred input methods where interaction did not require them to remove their hands from their handlebars, as this was perceived as a safer behavior. Traffic research confirms that single-handed control should be avoided, especially at low speed~\cite{schepers2012single}. Matviienko et al.~\cite{Matviienko:2018:ABH:3229434.3229479} proposed adding a multimodal warning signal to bicycles to increase awareness in critical cycling situations, with warning lights mounted at multiple points. Finally, BikeSafe~\cite{Gu:2017:BBB:3123024.3123158} showed that smartphones could be effectively used to detect dangerous cycling behavior and thus contribute to preventing dangerous traffic incidents. Past work illustrates the potential of interactive technologies to augment the safety of cycling, on which this paper attempts to build. We also observe that all the works cited here found that the way the cyclist moved with reference to the bicycle was of particular concern for interaction design. Our work investigates this aspect further.

Research has also explored design alternatives for input methods for in-ride interactions. Vechev et al.~\cite{Vechev:2018:MOA:3206505.3206527} proposed tapping different areas on the user's body to activate features while cycling. This strategy, however, requires removing one's hands from the handlebars, as does gestural input~\cite{Dancu:2015:GBE:2817721.2817748, Carton:2012:DCA:2370216.2370341}. S\"{o}r\"{o}s et al.~\cite{Soros:2013:CPB:2543651.2543660} suggested that voice navigation could be used to control an augmented reality display for cycling. Speech interaction, however, is sensitive to noise and a set of design limitations~\cite{Ghosh:2018:ETD:3173574.3173977}. Hochleitner et al.~\cite{Hochleitner:2017:NNS:3152832.3152871} designed smartphone controls for cycling and compared touch, button and wristband interaction. The focus of the study was designing game controls for outback cycling. In contrast, our work focuses on using smartphones in traffic and the impact of smartphone operation on bicycle control. Overall, past research offers little consensus on what the effective and safe ways of providing input during cycling are. As past work in traffic studies provide empirical evidence of what behaviours are unsafe~\cite{DEWAARD2014196}, the challenge for HCI is to design input methods that discourage those unsafe behaviours. Our work explores and compares in-ride input methods and examines their potential for safety and performance. We contribute the first study, to our knowledge, that empirically examines the impact of using a smartphone on performance in everyday tasks and bicycle control.

Augmenting the user experience of cycling has also been explored in past research. Rowland et al.~\cite{Rowland:2009:UCD:1613858.1613886} postulated including more context-aware interactions in cycling technology and embracing the qualitative experience of a bike ride. Further, they stressed how eyes-free audio control was important to a smooth cycling experience. Our work investigates the ways users can effectively control in-ride audio. A number of systems stressed the social experience of cycling and highlighted the role of the cyclist as part of a larger community of those in traffic. Biketastic~\cite{Reddy:2010:BSM:1753326.1753598} explored how routes could be annotated to estimate how friendly they are for cyclists. Ari~\cite{Andres:2019:MEG:3322276.3322307} was an e-bike designed to help users cross only on green lights. GameLight~\cite{Zhao:2019:GGO:3301019.3325151} attempted to transform cycling into a social exergame. Walmink et al.~\cite{Walmink:2014:DHR:2540930.2540970} built a system that displayed the rider's heart rate on the back of their helmet for social sharing. Our work explores a different dimension of the social aspect of cycling. We investigate the means for cyclists to display safe behaviours.

\subsection{Parallels to cars and motorbikes}
A bicycle is part of a modern traffic environment that also includes other vehicles. Consequently, work that explores input while driving cars or motorbikes is related to our inquiry.
Despite the differences that characterise the experience of driving each of these means of transportation, some research focuses on common aspects and useful insights can be derived for the particular case of bicycles.

The automobile interface field proposed several designs that keep the driver's hand on the steering wheel while providing second-task input. Modern cars with button-equipped steering wheels are a prime example. Research proposed alternative solutions such as multi-touch~\cite{Pfeiffer:2010:MES:1753846.1753984} or pressure-based input~\cite{Ng:2017:EIC:3025453.3025736}. Similarly, commercial motorcycles use buttons located next to the handgrips to control secondary functions~\cite{matre1982motorcycle}. These works not only provide inspiration for the design space we explored, but also showed that a number of complex tasks can be effectively and safely performed while controlling a vehicle. Our work explores this paradigm in a bicycle context.

\subsection{Input during Physical Activity}
As cycling is an outdoor physical activity, our work is also inspired by past efforts on how to use these methods during physical activity~\cite{Jones:2018:HOU:3170427.3170624}. Past work suggests that the number and complexity of available interactions under physical exertions (which cycling is likely to cause) should be limited. Wozniak et al.~\cite{Wozniak:2015:RRS:2785830.2785893} showed this requirement to be valid for running and Mencarini et al.~\cite{Mencarini:2016:DOW:2971485.2971509} for climbing. Their findings echo the constraints for interaction in motion as postulated by Marshall and Tennent~\cite{Marshall:2013:MIE:2468356.2468725}. In line with this work, our approach first identifies the most necessary inputs, considers motion constraints and investigates how users interact with their environment while providing input. We adopt a motion-centric design process, where we aim to minimise the complexity of the interactions while biking and design controls that affect the user's and the bicycle's movement to the least degree.

\section{Design}
Brotate and Tribike were developed in an iterative user-centred design process. Here, we provide an account of how we reached the final designs of our prototypes. Research in traffic safety shows that cyclists use mobile phones regularly~\cite{Ichikawa,TERZANO201387} and that this usage constitutes a risk to traffic safety~\cite{DEWAARD2014196}, but does not establish what functions of the phone are most commonly used. Consequently, we started designing a new control device for the smartphone by investigating what should be controlled. To that end, we conducted an online survey.
\subsection{Online survey}
We designed an online survey to investigate user requirements for interacting with a smartphone while cycling. Further, we investigated the user experience and social acceptability of in-ride mobile phone use. In the first part of the survey, we asked participants to rank the frequency of performing smartphone actions on a six-point Likert scale from `never' to `very often'. Then we asked how often and how they operated their phone while cycling: using their hand, headphone remote or dedicated Bluetooth device. In the second part of the survey, we presented the users with four actions performed while biking, in a randomised order: calling with the phone next to one ear, writing a text message with one hand, writing a text message with two hands and operating the headphone remote with one hand. As suggested by Williamson and Brewster~\cite{Rico:2010:UGM:1753326.1753458}, we presented alternative contexts in which the actions could be performed: a busy crossing where cyclists should give way, a crossing with traffic lights and a quiet street. We then applied the method used by Montero et al.~\cite{Montero:2010:YUS:1851600.1851647}, asking an open question: `What would you think if you saw someone else performing this gesture?' and asked the participants to rate their answer to: `How  would  you  feel about  performing  this  gesture  in  the  following  situations?' on a six-point Likert scale from `embarrassed'  to  `comfortable'. The survey design and the complete answer set are available as auxiliary material.  

\subsubsection{Participants}
We recruited 154 participants (119 male and 35 female, aged from 17 to 71, $M = 36.55$, $SD = 13.48$) via social media posts and snowball sampling. Forty participants identified as moderate cyclists, cycling 4 to 6 hours a week, and 55 reported cycling more than 6 hours a week. Participation was voluntary.
\subsubsection{Findings}
 Most participants (95 participants, 61\%) reported that they used their mobile phone on their bike. The majority of them operated the phone with one hand. Eleven users always controlled their phone using a Bluetooth remote control, while 61 out of 95 never used such a device. We used a two-way ANOVA on aligned-rank-transformed~\cite{Wobbrock:2011:ART:1978942.1978963} data to investigate the effect of \textsc{type of action} and \textsc{context} on the acceptability of the action. There was a significant main effect of \textsc{type of action} ($F_{3,1081}=303.89$, $p<.001$) and \textsc{context} ($F_{2,1092}=436.67$, $p<.001$). No interaction effect was observed, $F_{6,14}=1.03$, $p>.05$). Holm-corrected post-hoc comparisons revealed significant differences for all \textsc{type of action} and \textsc{context} pairs at the $p<.001$ level. Holding the phone in both hands was perceived as most unacceptable and the participants were most likely to accept in-ride smartphone use on quiet roads. Table~\ref{tab:survey} presents the results in detail. Two researchers used affinity diagrams to analyse the survey data from the open questions. Participants' comments focused on an inherent need for safety, changes from typical cycling posture, and responses to the traffic situation. We observed that users found many of the behaviors in the survey embarrassing and potentially dangerous. Low usage rates for Bluetooth controls showed that consumer-grade devices were not used widely. Finally, controlling music, answering calls and activating the voice assistant were identified as the top functions to be controlled. Interestingly, social media use was ranked as the least often used.

\begin{table}[]
    \centering
    \begin{tabular}{cccc}
    \toprule
         & busy crossing & traffic light & quiet street \\
         \midrule
         two-hands & \cellcolor{red!25}1.4 & \cellcolor{red!25}1.7 & \cellcolor{red!10}2.9 \\
         one-hand & \cellcolor{red!25}1.7 & \cellcolor{red!10}2.1 & \cellcolor{green!5}3.4 \\
         calling & \cellcolor{red!10}2.1 & \cellcolor{red!10}2.6 & \cellcolor{green!15}4.1 \\
         remote & \cellcolor{green!5}3.2 & \cellcolor{green!5}3.8 & \cellcolor{green!25}5.0 \\
         \bottomrule
    \end{tabular}
    \caption{Quantitative results of the survey. Acceptance was rated on a six-point Likert scale from 'embarrassed' to 'comfortable'  for four TYPES OF ACTION and three CONTEXTS. All pairs showed significant differences.}
    \label{tab:survey}
\end{table}

\subsection{Low-fidelity prototypes}
The next step in our design process was creating speculative prototypes of possible solutions. Having previously analysed relevant research work, we also analysed commercial solutions for inspiration. We reviewed commercially available solutions to find that devices featured five or more buttons arranged in a circular or linear pattern\footnote{See \url{https://buy.garmin.com/en-GB/GB/p/621230} or \url{https://cobi.bike/product}}. They were also all intended to be mounted in the middle of the handlebars. Consequently, we designed alternatives with less complexity that also minimised the hand movement required to operate the device.

\begin{figure}
    \centering
    \includegraphics[width=0.8\columnwidth]{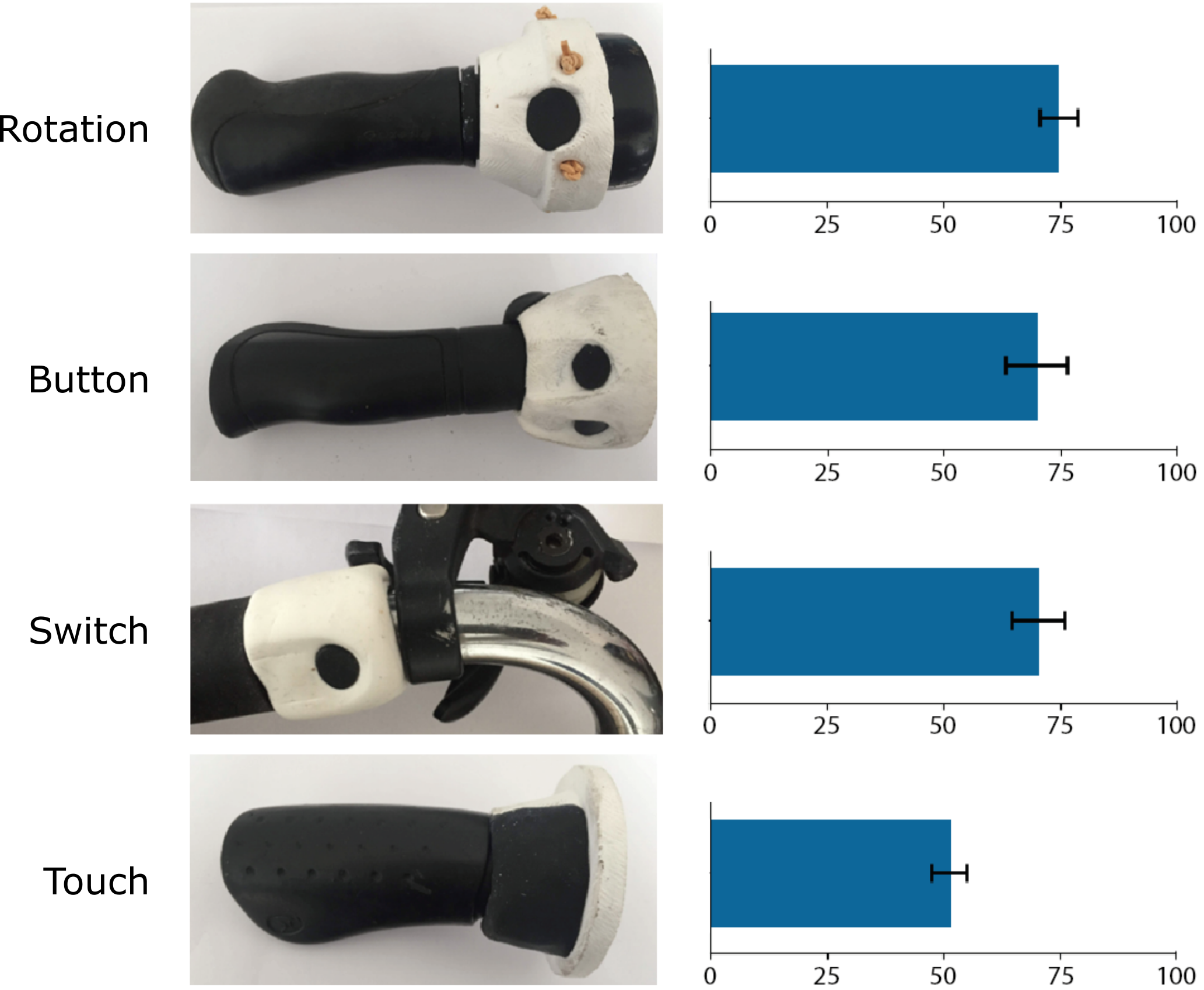}
        \Description[Low-fi protypes]{The four low-fidelity prototypes used in the design process. First the rotation device, second the button device, third the switch device, fourth the touch device. Below every photo is a bar chart showing the System Usability Score (SUS) for each device. Rotation has the ebst score.}
    \caption{The four low-fidelity prototypes used in our design process. Bar charts below the prototype pictures show SUS scores for each of the devices.}
    \label{fig:lowfi}
\end{figure}

We built four low-fidelity prototypes of possible smartphone control devices, presented in Figure~\ref{fig:lowfi}. First, the \emph{button} device, inspired by headphone remotes and motorbike controls, featured three buttons placed in the direct vicinity of the handlebar grip. Second, the \emph{touchpad} device, inspired by touch controls in modern cars, featured a touch-sensitive surface next to the grip. Third, the \emph{rotation} device, which mimicked rotational bells, offered input through rotating it around the handlebar axis and an additional button. Finally, the \emph{lever} device used derailleur control levers with an additional button. All prototypes were built using existing bicycle parts and moldable clay. We then conducted a study to gather user feedback about the prototypes.

{}

\subsubsection{Participants}
We recruited 14 participants (9 male and 5 female, aged 22--61, $M = 30.36$, $SD = 13.05$) through snowball sampling. All participants reported cycling regularly, i.e. more than 4 hours a week. They all stated that they used their mobile phones while cycling. 
\subsubsection{Procedure}
\begin{figure}
    \centering
    \includegraphics[width=\columnwidth]{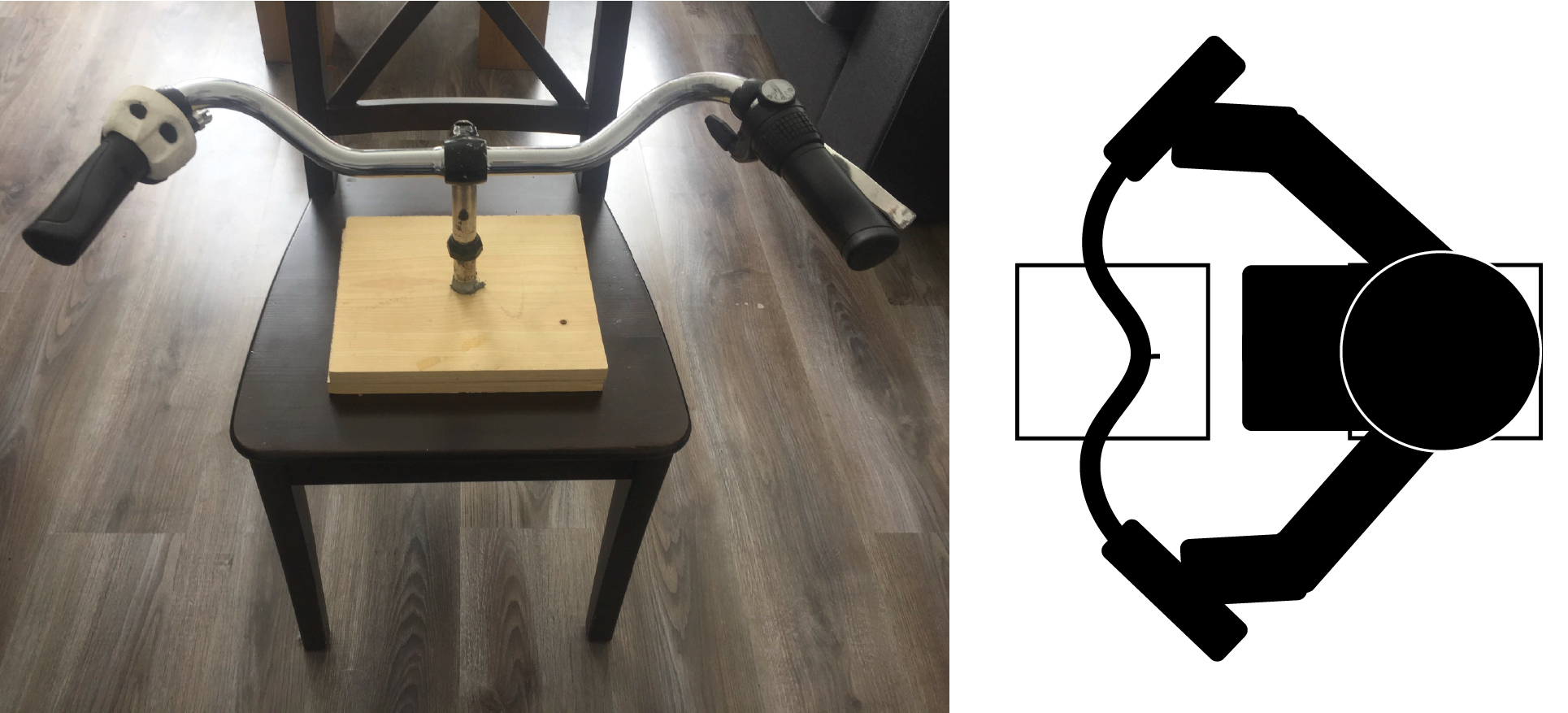}
    \Description[Pre-study]{The pre-study apparatus with the handlebar set on a chair in front of the participant.}
    \caption{The pre-study apparatus. The four low-fidelity prototypes were mounted on the handlebars to allow users to imagine using the devices while cycling.}
    \label{fig:setup}
\end{figure}

We mounted handlebars with a stem on a wooden plate to conduct our study. The four prototypes could be easily mounted and removed from the handlebars. The setup of this can be found in Figure \ref{fig:setup}. The participants were presented with each of the prototypes in a counter-balanced order. We used a think-aloud protocol where we asked the participants to imagine how they would use each of the devices to control their smartphone. After the participants interacted with the prototype, we administered the SUS scale to measure the anticipated usability of each of the devices. After all four devices were presented, we asked participants to rank them in order of preference. Finally, for the devices that incorporated motion, i.e. the \emph{rotation} and \emph{touchpad} devices, we conducted a gesture elicitation for the three most-requested smartphone functions---controlling calls, music and the voice assistant. The elicitation was conducted according to the procedure designed by Wobbrock et al.~\cite{Wobbrock:2009:UGS:1518701.1518866} with 9 referents, see Table~\ref{tab:task}. 

\subsubsection{Findings}
We used a one-way ANOVA on aligned-rank-trans\-formed~\cite{Wobbrock:2011:ART:1978942.1978963} data to compare SUS scores between the devices. There was a significant main effect of the type of device, $F_{3,39}=3.05$, $p<.05$. Post-hoc analysis revealed that the \emph{touchpad} device scored significantly lower than the other devices, all at $p<.05$. In terms of rankings, participants opinions were divided, with the button ranked best and the touch device being less preferred. The participants' remarks were transcribed verbatim. As in the online survey, two researchers analysed the data with affinity diagramming, focusing on the differences between the devices. Participants commented extensively on how the devices could be integrated into everyday commuting and how they affected using other bicycle controls. The \emph{lever} device was identified by most participants as potentially interfering with other actions and using a form reserved for derailleur control. Many participants also suggested using a longer grip and placing the devices further towards the centre of the handlebars, to assure a firmer grip for bicycle control. Elicitation results were classified and gesture sets were compiled using agreement scores as suggested by Wobbrock et al.~\cite{Wobbrock:2009:UGS:1518701.1518866}.

\subsection{Design requirements}
The survey showed that many users needed to operate their phones while cycling with a general negativity to use additional devices to do so. We gathered the insights gained in our design process so far using affinity diagramming and reviewed requirements in other work that investigated handlebar controls, e.g.~\cite{Kiss}. Based on that, we defined design requirements for future prototypes.  The smartphone control system should:
\begin{itemize}
    \item provide easy access to music control and answering calls. These were the top functions in our survey;
    \item enable the user to keep both of their hands on the handlebar. This requirement provides increased safety. It also reflects legal regulations in some countries.
    \item limit the hand movement on the handlebars. A steady hand position on the handlebars ensures minimal deviation from the straight line~\cite{DEWAARD2014196};
    \item not interfere with bicycle controls, e.g. brakes, derailleur levers, or the bell. This was a concern often mentioned by participants when reflecting on the low-fidelity prototypes;
    \item enable eyes-free operation as a secondary task, which was both a voiced user need and a safety requirement;
    \item convey an impression of safe use. The majority of the participants in the survey and low-fidelity prototype study commented extensively on the need for a bicycle device to evoke safety.
\end{itemize}

In light of these design requirements, we decided to further refine and evaluate the \emph{button} and \emph{rotation} devices. We eliminated the \emph{touchpad} device that users perceived as significantly less usable. We did not develop the \emph{lever} further, because of the participants' remarks on it interfering with bicycle controls. Building on that insight, we decided to place our devices on the left side of the handlebars as designed and artefact for commuters. City bikes usually feature less controls on the left side of the handlebar as they often have only one brake lever and zero or one derailleur controls.
\subsection{Final design}
Two devices---Brotate and Tribike were the final products of our design process. These are refined versions of the earlier \emph{button} and \emph{rotation} prototypes. Both devices were built with our design requirements in mind and in a form that considered the shape and function of the bicycle handlebars.

\subsubsection{Brotate}
\begin{figure}
    \centering
    \includegraphics[width=0.85\columnwidth]{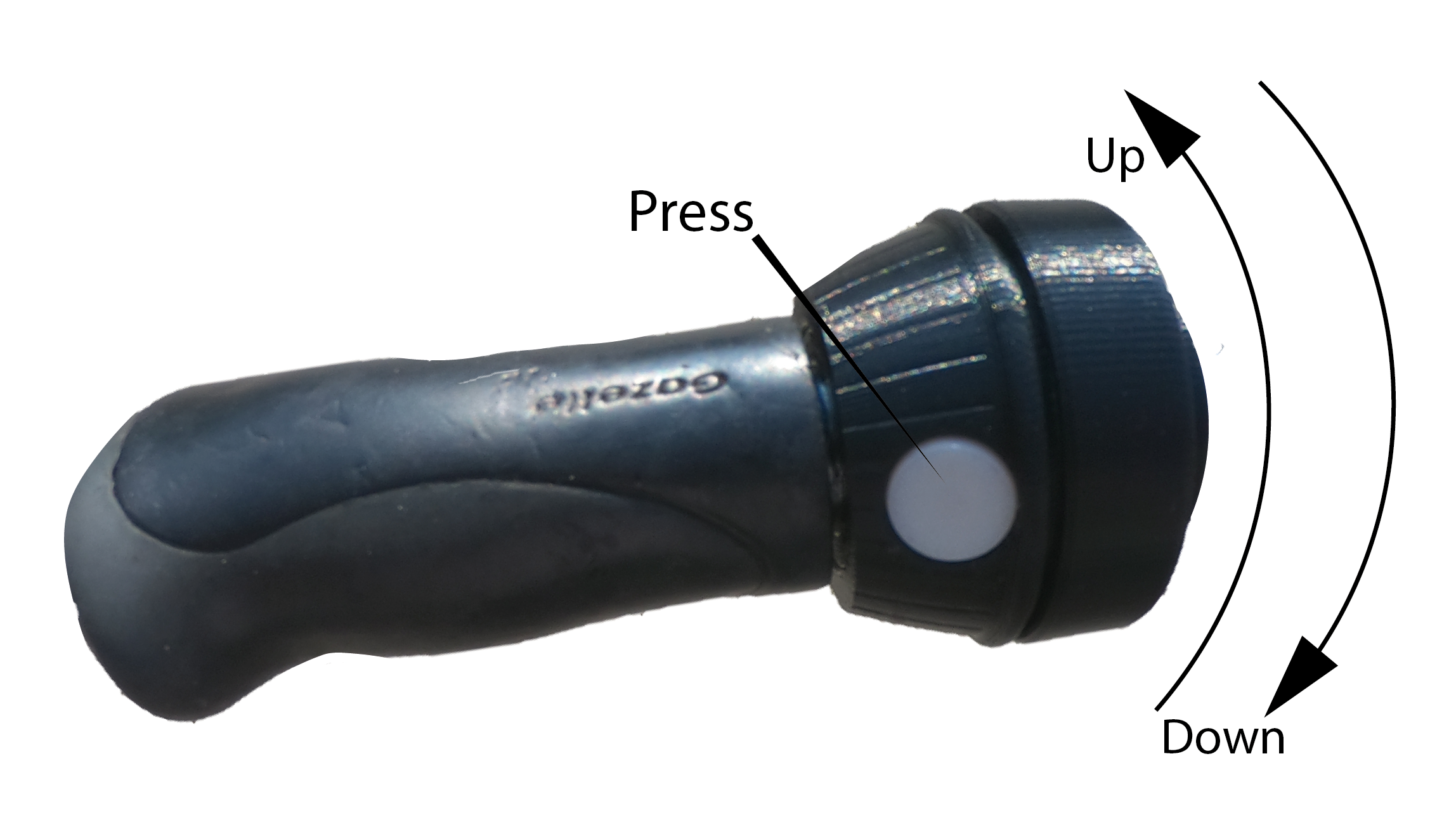}
        \Description[Brotate]{The Brotate device with its supported actions.}
    \caption{The final prototype of Brotate and its supported actions.}
    \label{fig:brotate}
\end{figure}

Brotate enables rotating part of the grip of the handlebars to control the cyclist's smartphone. The device supports two basic movements---forward and backward. To extend the range of functions supported, we added a single button located directly at the grip. Brotate is designed in a way that enables the cyclist's thumb to constantly rest on the button, thus requiring minimal movement for context-based actions. Figure~\ref{fig:brotate} shows the device and the range of supported motion. The control techniques used represent the gesture set from the elicitation process conducted using the low-fidelity prototype. Brotate provides tactile feedback through resistance when rotating. The rotating part returns to its starting position once an action is completed. Studies of the low-fidelity prototype revealed that a shorted grip was needed for Brotate so that the flesh of the palm could rest comfortably on the rotating part also allowing to easily move away from it.

{}

\subsubsection{Tribike}
\begin{figure}
    \centering
    \includegraphics[width=0.65\columnwidth]{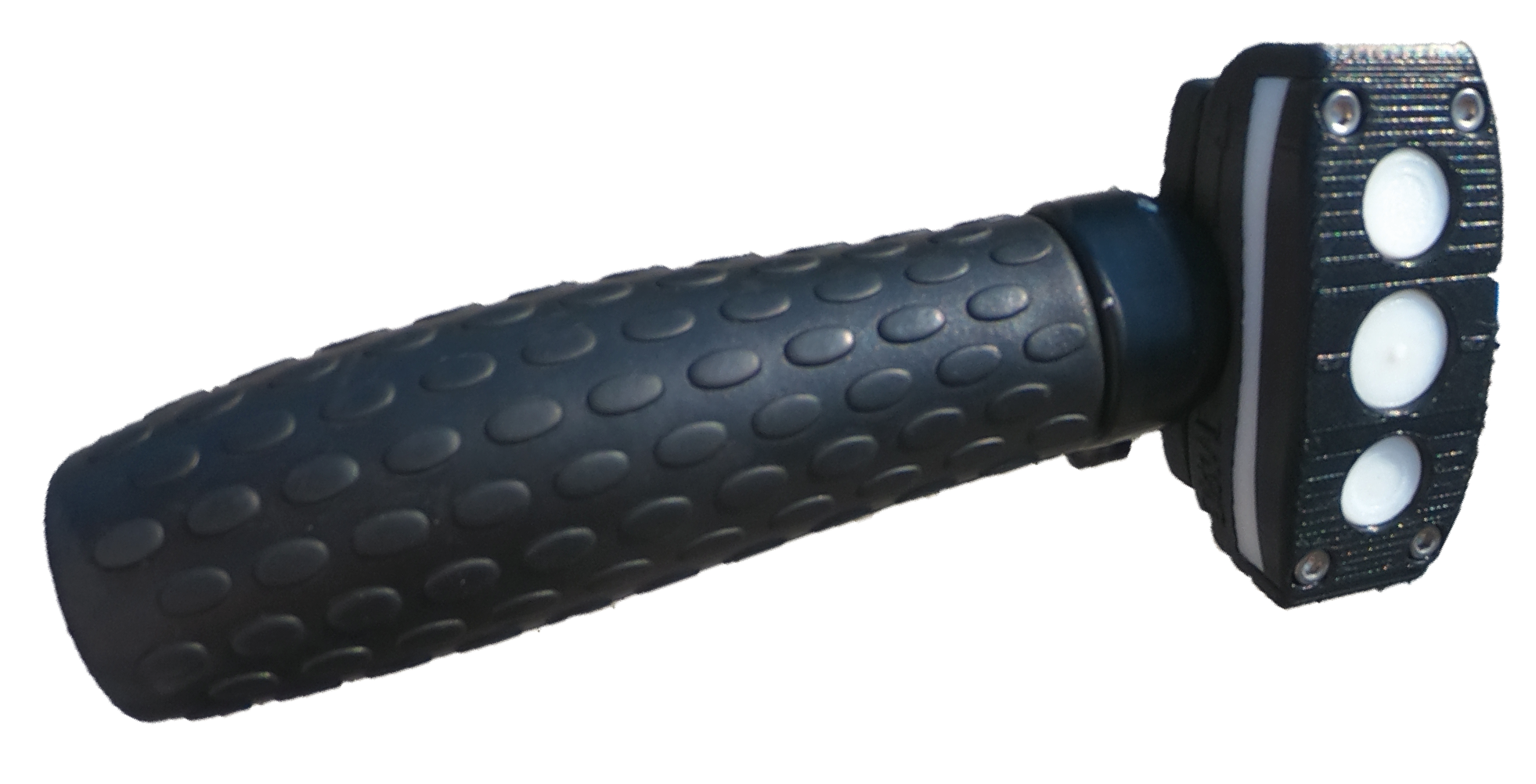}
    \Description[Tribike]{The Tribike device.}
    \caption{The final version of Tribike with three action buttons. The device uses the same button patterns as standard mobile phone in-line remotes.}
    \label{fig:tribike}
\end{figure}

Tribike is heavily inspired by headphone remotes, as these were positively perceived in the survey and the button device was highly ranked in the low-fidelity prototype study. Further, Tribike's form factor borrows from devices seen on motorbikes which benefit from the proximity of the buttons to where hands are usually placed. As suggested by users, Tribike uses a button wrap-around alignment where buttons can be reached with the thumb with little rotation of the palm holding the handlebars. The three buttons are highly tactile and feature additional spacing to counteract accidental presses. As the three horizontal buttons use the same layout as a standard in-line headphone remote, with which users are familiar, we used standard mobile phone button patterns to control the smartphone. Figure~\ref{fig:tribike} shows the final Tribike prototype.

\begin{table}
    \centering
    \begin{tabularx}{0.49\textwidth}{XXX}
    \toprule
        \textbf{Task} & \textbf{Tribike} & \textbf{Brotate}  \\
        \midrule
         Answer call & \text{MB} & \text{B}  \\
         Decline call & MB twice & B twice \\
         Volume up & TB & Rotate up\\
         Volume down & BB & Rotate down\\
         Next song & TB twice & B + rotate up \\
         Previous song & BB twice & B + rotate down\\
         Pause music & MB & B \\
         \bottomrule
    \end{tabularx}
    \caption{The smartphone actions used as tasks in our experiment with descriptions of how these tasks were performed with button presses and/or rotations using Brotate (MB---middle button, TB---top button, BB---bottom button) and Tribike (B---button). Call-related actions are only available when a call is active.}
    \label{tab:task}
\end{table}

\section{Implementation}
\begin{figure}
    \centering
    \includegraphics[width=\columnwidth]{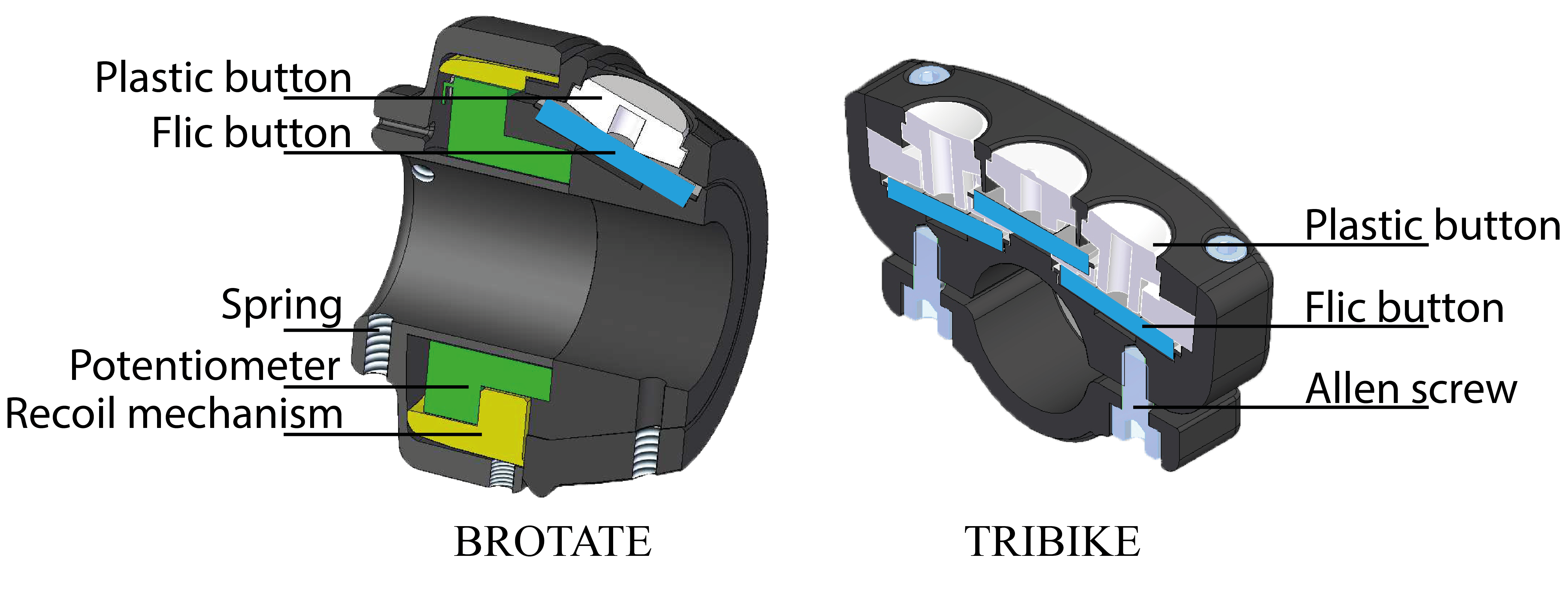}
    \Description[3D models]{3D models of the Brotate and Tribike devices.}
    \caption{Final 3D models of the Brotate (left) and Tribike (right). The spring inside Brotate, visible in the cross-section, provides tactile feedback when rotating the device. The recoil mechanism in Brotate includes a steel spring.}
    \label{fig:models}
\end{figure}

We built 3D models of both of the devices and 3D-printed the required components, as illustrated in Figure~\ref{fig:models}. Brotate uses a design where the outer part of the grip moves a Panasonic EWV-YG9U04B14 rotary potentiometer as it rotates. The base of the potentiometer is fixed to the handlebars. To provide tactile feedback and assure that the grip would return to its original position, we mounted a spring inside the device. The tactile buttons of both devices use flic\footnote{\url{https://flic.io/}} programmable buttons, which feature embedded Bluetooth Low Energy (BLE) connectivity. Rotary input in Brotate is processed by a DFRobot Beetle BLE Arduino-compatible microcontroller. We also built a custom Android application which logged all the device events for analysis. The devices connected to the user's smartphone using BLE and the inputs were mapped to smartphone actions using the Tasker\footnote{\url{https://play.google.com/store/apps/details?id=net.dinglisch.android.taskerm}} tool.

\section{Evaluation}
To evaluate Brotate and Tribike, we conducted a within-subjects experiment where we asked participants to complete smartphone-related tasks while cycling, using the two devices and controlling the phone with one hand. We endeavored to investigate how effective Brotate and Tribike were in controlling a smartphone while cycling. Further, we studied the usability of the devices and their impact on the bicycle's lateral control. Our work closely follows the study conducted by De Waard et al. ~\cite{DEWAARD2014196}. As our work looks specifically at HCI for physical activity and not traffic safety, we chose to adapt a study from the traffic safety domain rather than designing another protocol which would require further validation.

\subsection{Participants}
We used social media and flyers to recruit 19 participants (9 male and 10 female, aged 22--66, $M = 31.79$, $SD = 14.09$). All participants declared cycling regularly and 15 of them reported using their phone regularly while cycling. All participants had a history of multiple years of everyday commutes by bike, using equipment that required braking and derailleur control with devices mounted on the handlebars. Six participants were students and the rest were employees in governmental and private organisations. We provided an online shop voucher for \$10 as remuneration for their time, and free refreshments were available throughout the study. Prior to participating in the study, participants were asked to declare they were fit to cycle for 30 minutes at a moderate pace.

\subsection{Apparatus}
Two equivalent unisex city bicycles were used in the experiment. One was fit with Brotate and the other with Tribike. Both prototypes were fit next to the left handlebar grip of the respective bicycle. We chose the left side of the handlebar as consumer-market bicycles usually feature derailleur controls placed on the right side. Participants were given a Huawei P30 Lite smartphone running Android 9.0 with the study software installed and wired headphones. The experiment was recorded with a stationary camera and an action camera mounted on the bicycle that recorded all device interactions. An additional smartphone (Samsung Galaxy S8) was mounted on the seat post to measure acceleration throughout the experiment. We also mounted one-way radios centrally on the frames of the bicycles so that the experimenters could communicate with the participant at all times.

\subsection{Task and Procedure}
\begin{figure*}
    \centering
    \includegraphics[width=2\columnwidth]{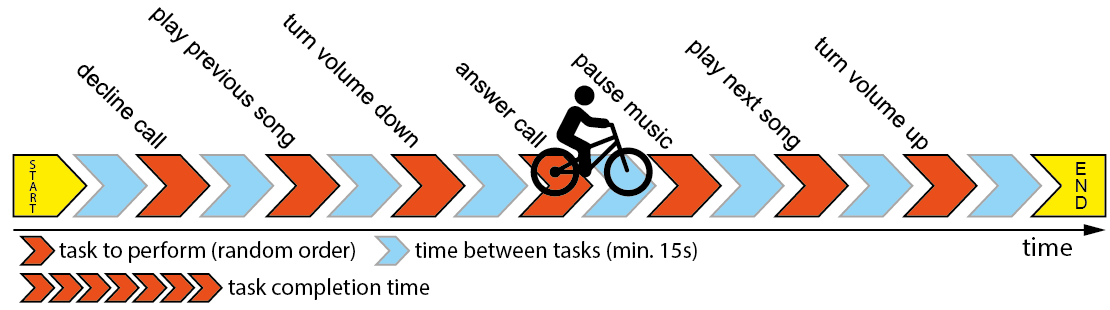}
      \Description[Study]{Schematic overview of an example sequence of tasks the participants had to perform in the experiment.}
    \caption{An example sequence of tasks which the participants performed in the experiment. Note that the tasks were only constrained by minimum time between them. Distance was determined by the participant's preferred moderate cycling speed. Refer to Table~\ref{tab:task} for the respective controller actions. Conditions were administered in a Latin square order while the order of the seven tasks was randomised.}
    \label{fig:task}
\end{figure*}

Our study focused on evaluating how the users could leverage Tribike and Brotate to operate the smartphones while cycling. To that end, we asked the participants to complete seven basic smartphone actions based on our survey: turn the volume up, turn the volume down, play the next song, play the previous song, pause the music, decline a call and answer a call. We used Latin squares to avoid order bias in administering the tasks. The task was communicated to the user with pre-recorded voice instructions, e.g. `Please switch to the next song now.' with the exception of the calls to be received or declined where the calling person's name indicated whether the participant was to decline or receive the call (the smartphone was set to read caller names for incoming calls). When a call was received, an experimenter thanked the participant for completing the task and disconnected. Throughout the experiment, participants were asked to cycle straight at their preferred, moderate speed on 400-metre long straight track. The track was located in a location with limited visual distraction, featuring only grass and trees on its entire length. We assured that the individual tasks were distributed in a way that there was a minimum cycling time of 15 seconds between completing one task and starting the next. This assured that participants could focus solely on cycling before attempting each task. Additionally, we ensured that no task was performed while the participant was turning around or within 10 seconds before or after the turn. Figure~\ref{fig:task} shows an example of a task sequence in the experiment. The order of the tasks was different within each trial so that the participants could not anticipate the next action to be performed.

After the participants arrived at the cycling track, we greeted them, explained the purpose of the experiment, and asked them to complete a demographics, data processing and consent form. Afterwards, we asked the participants to sit on their bicycles and adjusted the seat height as desired. We then assigned the initial condition to the participant, based on Latin-square order balancing. Next, for each condition, the participants completed a practice task in which they could review all the tasks and actions while stationary. We presented all the tasks in a fixed order and presented the required input. The participants were then free to use the device and complete tasks until they reported that they could comfortably use it while cycling. We then asked them to begin cycling, reminding them to cycle at a moderate and comfortable speed, in a line as straight as possible. The participants then completed the sequence of smartphone tasks as shown in Figure~\ref{fig:task}. After the tasks were completed in each condition, the participants completed NASA TLX and SUS questionnaires. The tasks were then completed for the remaining conditions with rest time as required in between. After completing all tasks in the study, we conducted a semi-structured interview with each participant. The interview focused on the perceived differences between the conditions, how using the devices affected the ride experience and the usability of the devices. After the conclusion of the experiment, we provided the participants with the remuneration and offered refreshments.

\subsection{Hypotheses}
Our experiment evaluated the following hypotheses, based assumption that Brotate and Tribike would alleviate the issues discovered by De Waard et al.~\cite{deWaard2}:
\begin{enumerate}
    \item Using Brotate and Tribike will reduce sideways movement while cycling compared to one-handed smartphone operation. We investigated if our designs could limit the negative impact on lateral control caused by one-handed smartphone use while cycling~\cite{DEWAARD2014196}.
    \item Brotate and Tribike will enable performing smartphone tasks more efficiently than one-handed control.
    \item Brotate and Tribike will be perceived as more usable and requiring less cognitive effort to use than one-handed control.
\end{enumerate}

\subsection{Conditions and Measures}
\textit{Device used} was the only condition in our study. The levels were: \textsc{One Hand}, \textsc{Brotate} and \textsc{Tribike}. We measured the total \textit{Task Completion Time (TCT)} for the seven tasks as the sum of the times for the individual tasks as a measure of effectiveness in control. The time was measured from the end of the voice command to the smartphone registering the user action. In the case of declining/receiving calls, the time was measured from the moment the participant's phone started ringing. The participant's phone also recorded the \textit{Error Rate}. Performing an incorrect action in a task was counted as an error. Raw \textit{NASA TLX} for and \textit{SUS} scores were collected using printed questionnaires at the conclusion of each condition. We used \textit{NASA TLX} as a cognitive load measure as it was linked to smartphone use while cycling by De Waard et al.~\cite{DEWAARD2014196}. \textit{SUS} was used to measure the perceived usability of the system.
% \begin{figure}[ht]
%     \centering
%     \includegraphics[width=0.4\columnwidth]{figures/acceleration.png}
%     \caption{The three acceleration vectors acting on the bicycle during the study. We used the raw values of the vector to estimate the tilt of the bicycle during the experiment.}
%     \label{fig:vector}
% \end{figure}

\begin{figure*}
    \centering
    \includegraphics[width=1.85\columnwidth]{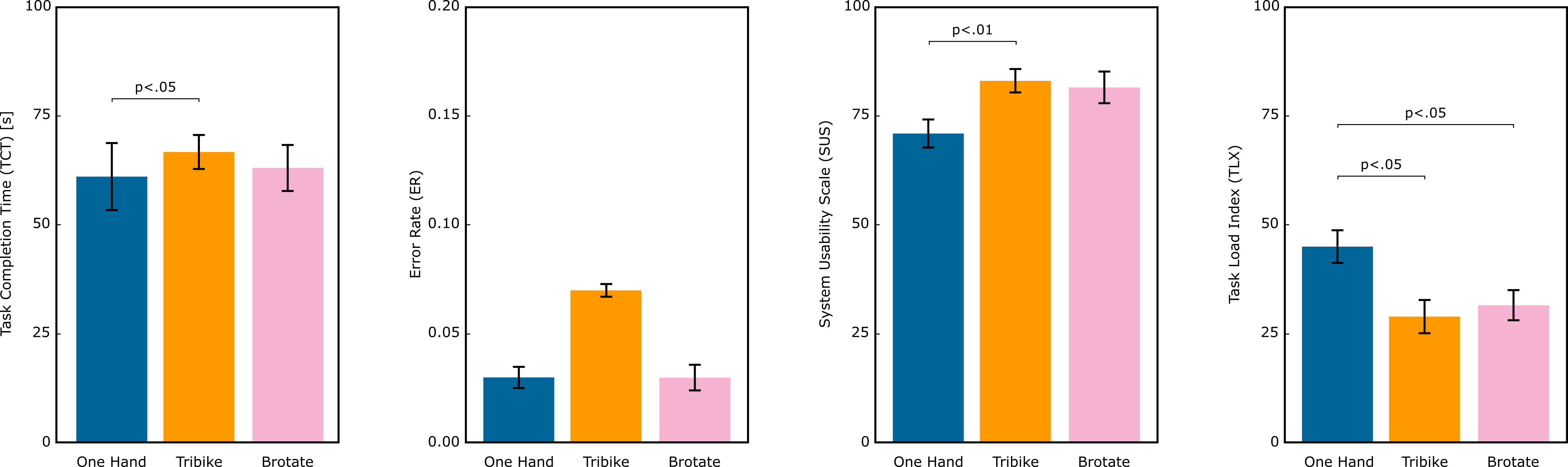}
    \Description[Results]{Visualisations of the results presented in Table 3.}
    \caption{Mean Task Completion Time (left) and Error Rates (centre left) in the experiment for the three experimental conditions. Error bars visualise standard error.  Normalised Mean Bike tilt values collected in our experiment. Error bars show standard error. The values were normalised to account for between-participant differences in riding style. Mean SUS (centre right) and Raw NASA TLX (right) scores collected in our experiment for the experimental conditions. Error bars show standard error.}
    \label{fig:results}
\end{figure*}

\begin{figure}[ht]
    \centering
    \includegraphics[width=0.85\columnwidth]{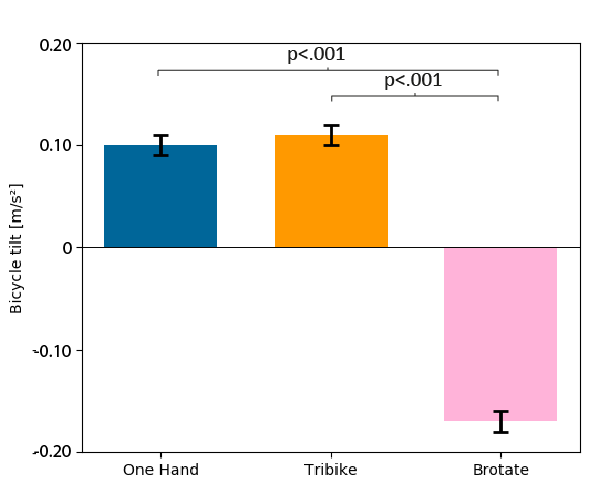}
    \caption{Normalised Mean Bike tilt values collected in our experiment. Error bars show standard error. The values were normalised to account for between-participant differences in riding style.}
    \label{fig:biketilt}
\end{figure}

Finally, the smartphone mounted under the bicycle seat recorded \textit{Bicycle Tilt}. De Waard et al.~\cite{DEWAARD2014196} empirically related lateral stability when cycling in a straight line to traffic safety. We used the inertial measurement unit (IMU) of the phone to capture the sideways movement of the bicycle through the experiment thus measuring lateral control. This approach was inspired by past work~\cite{joo2018new} that successfully used IMUs to measure bicycle motion. This is an alternative approach to De Waard et al.~\cite{DEWAARD2014196}, which we used because we aimed to study the effect of the different devices on bicycle movement, and not the nature of the movement itself. Our analysis investigates the amount of sway and/or tilt caused by using the smartphone control devices while cycling thus providing a measure of lateral control. We first applied a filter to the IMU signal to remove data from instances when the bicycle was stationary and when the participants were turning around. Possibly dangerous changes in the position of the bicycle would result in changes in the acceleration vectors acting on the bicycle. However, gravity and the acceleration caused by the participant pedalling also contribute to the overall acceleration. To factor out these vectors from our analysis and focus solely in sideways acceleration change, we defined a $100ms$ sliding window and integrated the signal for the three instantaneous acceleration values $a_{x}$ and $a_{y}$ as follows ($a_{z}$ represents gravitational pull and is constant):

\[ \int_{t}^{t+100ms} \sqrt{{a_{x}}^2+{a_{y}}^2} \quad dt \]

Finally, in order to account for each participant's individual comfortable speed, body movements and cycling style, the scores were standardised. Data was standardised using R’s normalize function. The initial acceleration period was removed from the data through matching with video timestamps.

\subsection{Results}
We used a one-way ANOVA to investigate the effect of the device used on TCT, error rate, NASA TLX scores and Bicycle Tilt. Significant main effects were observed for all measures but the Error Rate. Table~\ref{tab:results} presents the details of the analysis and Figure~\ref{fig:results} illustrates the results.

\begin{table*}
\begin{tabularx}{\textwidth}{Xd{4.4}d{4.4}d{4.4}d{4.4}d{4.4}d{4.4}d{4.4}d{4.4}}
\toprule
      & \multicolumn{2}{c}{\textbf{TCT [$s$]}}    & \multicolumn{2}{c}{\textbf{Error Rate}}  & \multicolumn{2}{c}{\textbf{NASA TLX}}    & \multicolumn{2}{c}{\textbf{Bicycle tilt [$m/s^2$]}}   \\
      
       \cmidrule(r){2-3}\cmidrule{4-5}\cmidrule(l){6-7}\cmidrule(l){8-9}
\textbf{Condition}       & \multicolumn{1}{c}{$M$} & \multicolumn{1}{c}{$SD$} & \multicolumn{1}{c}{$M$} & \multicolumn{1}{c}{$SD$} & \multicolumn{1}{c}{$M$} & \multicolumn{1}{c}{$SD$} & \multicolumn{1}{c}{$M$} & \multicolumn{1}{c}{$SD$} \\
       \midrule
\textsc{One Hand} & 61.11* & 7.67 & 0.03 & 0.00 & 45.44*\dagger & 16.96 & 0.10* & 1.06\\
\textsc{Tribike} & 66.63* & 4.06 & 0.07  & 0.01 & 29.33\dagger & 16.73& 0.11\dagger & 1.10\\
\textsc{Brotate} &  63.21 & 5.05 & 0.03 & 0.00 & 31.88* & 15.16 & -0.17*\dagger & 0.82 \\
\midrule
\textbf{ANOVA} & \multicolumn{2}{c}{ $F_{2,54}=4.36$, $p<0.05$}    & \multicolumn{2}{c}{$F_{2,54}=1.86$, $p=0.17$}  & \multicolumn{2}{c}{$F_{2,54}=5.04$, $p<.05$}& \multicolumn{2}{c}{$F_{2,57725}=524.1$, $p<.001$} \\
\bottomrule
\end{tabularx}
\caption{Mean values and standard deviations for the TCT, Error Rate, NASA Task-Load Index and Bicycle tilt with from respective one-way ANOVAs. * and $\dagger$ show significantly different pairs from post-hoc testing using Tukey HSD, at the~$p<.05$~level. Note that Bicycle tilt values were integrated and standardised, with normality requirements fulfilled (Anderson-Darling test, $A=3470.8, p < .001$).}
\label{tab:results}
\end{table*}
To further confirm the bicycle motion analysis, we ran an additional analysis Root Mean Square (RMS) $a_y$ values, i.e. the raw sideways tilt of the bicycle. Hand control resulted in the largest RMS $a_y$ values recorded, while TriBike produced the lowest values. An one-way ANOVA showed that there was a significant effect of \textit{device used} on RMS $a_y$: F(2,58723)=78.9, p<.001. Tukey HSD revealed that all condition pairs were significantly different at p<.01. These results are in line with our derived measure of tilt.

We applied the align rank transformation~\cite{Wobbrock:2011:ART:1978942.1978963} on SUS data to analyse it with a one-way ANOVA. Mean cell frequencies did not exceed 10~\cite{luepsen_aligned_2017}. There was a significant effect of device used on the SUS score, $F_{2,54}=6.65, p<.01$. Post-hoc test with Tukey HSD showed that there was a significant difference between \textsc{One Hand} and \textsc{Tribike}, $p<.01$. Figure~\ref{fig:results} shows detailed results.

The volume of recorded qualitative data was 200 minutes. Given the size of the data set, we adopted a pragmatic theme-based approach to data analysis as suggested by Blandford et al.~\cite{blandford2016qualitative}. Recordings of debriefing interviews were transcribed. Two researchers first identified relevant passages from the data, which were then open-coded independently by the two researchers. Based on the codes and iterative discussion, we identified three themes in the data: \textsc{integration} (with other bicycle components), \textsc{learning} (how to use the device), and \textsc{movement}.

\subsubsection{Integration}
The participants commented extensively on how Tribike and Brotate could be placed in ways that do not conflict with existing bicycle controls. Brotate was perceived as using existing bicycle control metaphors, such as the rotating bell: 
\begin{quoting}
The device [Brotate] is quite similar to a rotating bell, which is well integrated into your handlebars.
\end{quoting}
Further, the participants reported that they appreciated the fact that Brotate integrated into the handlebars and did not add another visible device to the bike controls. One participant contrasted the two devices:
\begin{quoting}
While it's [Brotate] a bit more complex than the button device, I like the fact that it's part of the handlebars.
\end{quoting}
In contrast, three of the users in the study were skeptical of integrating new devices into bicycle controls. One participant noted that their core motivation behind not using a device for smartphone control was the fact that he was already proficient with operating the smartphone using his hand:
\begin{quoting}
I am used to operating my phone in one hand on a bike and looking at it, so that's what I prefer.
\end{quoting}

\subsubsection{Learning}
Anticipating how users may gain proficiency in using Tribike and Brotate and considering the use of the devices over a longer time period was a strong theme in the interviews. Given the novelty of the devices, the participants reported that the devices appeared complex and training was needed:
\begin{quoting}
I'm not used to operating such a device [Tribike], so, in the beginning, it felt a bit strange.
\end{quoting}
However, no users mentioned having trouble controlling smartphone actions with Tribike or Brotate after completing the practice task and completing the experiment. Participants reported having familiarised themselves with the operation of the devices and being able to issue smartphone commands without looking:
\begin{quoting}
I could operate the device [Brotate] without looking, especially now, when I'm more familiar with the functions.
\end{quoting}
Relating Tribike and Brotate played a role in how quickly users learned how to use the devices. Tribike's use of the headphone remote layout was helpful to users. Consequently, Brotate was perceived as more novel and required more time for acquainting:
\begin{quoting}
Pressing the buttons feels more familiar than rotating the device.
\end{quoting}

\vspace{-0.7em}
\subsubsection{Movement}
This theme describes the users' perception of the hand movements required to operate one's smartphone when cycling and how these movements impacted their performance in the tasks. Users felt that a key advantage of Brotate was the lack of a need to re-position one's hand to operate it:
\begin{quoting}
I didn't have to change my grip on the handlebar when using it. [Brotate]
\end{quoting}
In contrast, participants also remarked that wrist rotation was not a movement they associated with riding a bicycle. Some users reflected that Brotate required movements that were suboptimal:
\begin{quoting}
To make the movement [Rotate up], you need to slightly rotate your wrist, which felt unnatural.
\end{quoting}
Finally, some users found it difficult to operate the devices due to the size of their hands and palms. One user reported that he needed extra movement to hit the buttons on Tribike because of his large hands:
\begin{quoting}
I needed to adjust my grip to operate the device [Tribike], maybe because my hands are relatively big. 
\end{quoting}

\section{Discussion}
Throughout our design process and evaluation of Tribike and Brotate, we observed that the devices benefited interaction while cycling. Here, we summarise our findings and provide suggestions for future systems which aim to support cyclists.

\textbf{One-handed smartphone use offers limited performance benefits at the cost of cognitive load.}
Our experiment revealed that the participants were able to complete the tasks with the three devices with equal accuracy. One-handed smartphone use was, however, faster than the other methods, which implies that \textit{H2} was not confirmed. As the majority of the study participants did use their phones while cycling regularly using the one-handed method, we believe that this result can be partially attributed to their acquired proficiency. This was also confirmed by qualitative data in the \textsc{learning} theme. Further, the results show a significant increase in cognitive load when operating the phone with one hand. This result is congruent with the findings by De Waard et al.~\cite{DEWAARD2014196}. This can be explained by the fact that Tribike and Brotate do not require visual attention. While the phone can provide richer visual feedback, it appears that it does not aid the users in basic actions. Further, past work has shown that strong haptic feedback is particularly useful when designing for dual tasks~\cite{Burke:2006:CEV:1180995.1181017} such as cycling and using a phone. These findings also echo results from past studies in interaction during physical activity which also showed a need for highly tactile controls, e.g. in the context of climbing~\cite{Spelmezan:2009:TMI:1518701.1519044} or running~\cite{Curmi:2017:ECI:3025453.3025551,Wozniak:2015:RRS:2785830.2785893}.

SUS scale scores from our experiment provide additional motivation for developing alternative means of controlling the smartphone while cycling---users ranked issuing commands by hand significantly lower despite the majority of the being proficient in one-handed operation. This is in contrast with the NASA TLX scores. Thus, \textit{H3} is partially confirmed. In contrast with past work~\cite{Hochleitner:2017:NNS:3152832.3152871}, our study provides empirical evidence that handlebar-mounted smartphone controls can effectively reduce the cognitive load required to perform smartphone actions while cycling. This fact implies that using bicycle-specific controllers limits the attention required for the interaction and leaves more cognitive capacity for traffic. Consequently, future bicycles should offer the means for eyes-free interaction with the cyclist's smartphone.

\textbf{Using Brotate limits undesirable bicycle movements.}
The IMU measurements from our experiment showed that Brotate offered superior performance in terms of lateral control of the bicycle through the ride compared to the other conditions. This implies that \textit{H1} is confirmed for Brotate. Such a result suggests that the bicycle was more stable when using Brotate and, consequently, the cyclist had more control over the riding path. While De Waard et al.~\cite{DEWAARD2014196} also observed excessive swaying when the phone was held in one hand, primarily caused by an altered body position on the seat (operating a smartphone with one hand caused cyclists to sit more upright), the significant difference between Brotate and Tribike should be attributed to other causes. Our design process assured that both devices were located as close as possible to the handlebar grips and did not require excessive movement to activate. The difference between the two devices can be explained not by the range of movement required to perform the actions using Brotate and Tribike, but the direction of the motion. In the case of Brotate, the rotation move is performed parallel to the axis of motion of the bicycle. In contrast, placing the thumb on the top or bottom buttons on Tribike requires movement in a place perpendicular to the direction in which one is pedalling. As a consequence, our work implies that future devices for providing input while cycling should minimise the amount of movement along the handlebars. The study results for Brotate suggest that techniques that use handlebar grips offer most lateral control. Further, understanding the motion required to operate the device in the context of the motion of the bicycle is a key design consideration.

\textbf{New devices for cycling should leverage existing form factors and user-specific configurations.}
In our design process, we built devices that were explicitly inspired by existing bicycle controls. Results in the \textsc{integration} theme show that users appreciated the familiar form factors of a bell and headphone remote. Furthermore, we ensured that other controls of the bicycle were not affected by Brotate and Tribike. The \textsc{movement} theme showed that the smartphone controller not only needed to integrate well with existing controls, but it should also be subject to the same user-specific requirements as other bicycle components. Palm size is a known limitation when designing for bicycle controls, which can be observed in commercial products such as integrated derailleur-brake controls~\cite{yamashita2000brake}. This finding fits within a larger trend of augmenting existing equipment in building technology for exertion~\cite{clair,shoe}.

While this observation sets design constraints for new devices for cycling, it also offers opportunities. Firstly, as bicycle technology evolves, cycling-related controls require less space and force to operate. For example, electronically controlled gear shifting no longer requires the activation force of older, cable-based systems, and  eBikes~\cite{Andres:2019:MEG:3322276.3322307} often feature automatic gear shifting. Consequently, the number of controls required for cycling and their form factor is reduced. Our study illustrates that interactive devices for cycling can effectively re-purpose existing control metaphors for purposes like smartphone control. Future systems for cycling should prioritise existing handlebar-based controls over novel methods to effectively use the users familiarity with those controls and prevent undesired operation.

\subsection{Limitations}
As our work constitutes an exploratory inquiry into smartphone control while cycling, we are aware of certain compromises and limitations to which our research is prone. We decided to not investigate controlling smartphones mounted in a holder on the handlebars. This decision was motivated by the fact that we wanted to focus on devices that limited the required visual attention. Additionally, past work in the automotive domain showed detrimental effects of holders~\cite{Kujala:2015:HTL:2818187.2818270}. However, we do recognise that smartphones stored in holders may soon appear on the streets more frequently and may need to be studied. Future research should compare eyes-free devices like Brotate and Tribike with holder-based solutions in terms of performance and safety. We also note that none of the participants reported ever using the smartphone buttons while not looking at the phone, e.g. in one’s pocket, which would be a way to use tactile controls. While we hypothesise that such a solution would be safer than looking at the phone, it still requires moving a hand away from the handlebars thus negatively affecting stability.

Further, we recognise that our results may be affected by the fact that most participants of the final study were proficient cyclists who reported regularly using a smartphone while biking. We opted to design for this user group, because we endeavored to understand the design possibilities for more efficient and potentially safer controls for them. However, the rising population of cyclists implies that new users are constantly introduced to smartphone usage while cycling. Future research should investigate designing devices for novice bicycle commuters who may have different needs. We also note that the actions we designed for were determined in our survey. However, we need to consider that a declared preference for calls and music control might be due to social desirability bias. Further, our user-centred design process was conducted with current users of city bikes and thus the design of our devices is limited to those users. More studies of behaviour of different types of cyclists are needed in traffic research to better inform design. 

Finally, we believe that a broader discussion is needed about the social implications of our design. While increasing safety through reducing the number of users holding their phones while cycling is one of our primary motivations, there is a possibility that light-weight eyes free smartphone controls may increase the overall number of cyclists using smartphones. Certain social trade-offs are involved with cycling technology, such as the acceptability of such behaviors and using protection devices~\cite{deWaard2}, and the balance of such should be monitored. We hope that social science studies in traffic can benefit from our understanding of interacting with technology while cycling and effectively affect social acceptance and policy decisions. Further, smartphone use \emph{per se} does have a negative impact on safety~\cite{deWaard2}, and it remains a challenge to manage the cyclists' attention to assure comfort and safety. An emerging question is how to design technology that would enable effective and safer smartphone use while cycling and, simultaneously, not encourage smartphone interactions when they are not strictly necessary.

\section{Conclusion}
In this paper, we presented the design, implementation and evaluation of two smartphone controllers for cyclists---Brotate and Tribike. We first conducted a user survey to identify the most common smartphone actions used while cycling. We designed Brotate---a device that uses handlebars grip rotation and Tribike---a handlebar-mounted device with three buttons in an iterative process. In an experiment, we compared the two devices with one-handed smartphone control. The study showed that both devices enabled efficient control while significantly reducing cognitive load. Further, Brotate significantly improved lateral control during cycling. Our work constitutes a structured exploration of smartphone input while cycling. Our results show that future designers should focus on understanding and limiting movement across the handlebars, leveraging existing interaction metaphors for bicycles and providing user-specific solutions. We hope that our work inspires further inquiries into input while cycling, which can contribute to increasing traffic safety.

% BALANCE COLUMNS
\balance{}

% REFERENCES FORMAT
% References must be the same font size as other body text.
\bibliographystyle{ACM-Reference-Format}
\bibliography{sample-base}

\end{document}